\documentclass{article}
\usepackage{float}
\usepackage{amsthm,,amssymb, color, epsfig, graphics}
\usepackage{graphicx}

\usepackage[intlimits]{amsmath}
\usepackage{latexsym}

\usepackage{tikz}
\usepackage{tikz-3dplot}
\usetikzlibrary{positioning,calc}

\def \h#1{\widehat{#1}}
\def \t#1{\widetilde{#1}}
\def \xht#1{\widehat{\widetilde{#1}}}
\def \b#1{\overline{#1}}
\def \xbt#1{\overline{\widetilde{#1}}}
\def \xbh#1{\overline{\widehat{#1}}}
\def \xbht#1{\overline{\widehat{\widetilde{#1}}}}
\newcommand{\bse}{\begin{subequations}}
\newcommand{\ese}{\end{subequations}}
\newcommand{\nn}{{\nonumber}}
\newcommand{\BA}{{B\"acklund\ }}
\def \pual {{{\vphantom{\frac{1}{2}}}}}

\title{Elementary introduction to discrete soliton
  equations\footnote{to appear in ``Nonlinear Systems and
    Their Remarkable Mathematical Structures'', Norbert Euler (ed.),
    CRC Press.}}

\author{Jarmo Hietarinta\footnote{e-mail: jarmo.hietarinta@utu.fi}
\\Department of Physics and Astronomy
 \\University of Turku\\FI-20014, Turku, Finland}

\begin{document}

\date{}

\maketitle

\begin{abstract}
  \noindent
  We will give a short introduction to discrete or lattice soliton
  equations, with the particular example of the Korteweg-de\ Vries as
  illustration.  We will discuss briefly how B\"acklund
  transformations lead to equations that can be interpreted as
  discrete equations on a $\mathbb Z^2$ lattice.  Hierarchies of
  equations and commuting flows are shown to be related to
  multidimensionality in the lattice context, and multidimensional
  consistency is one of the necessary conditions for
  integrability. The multidimensional setting also allows one to
  construct a Lax pair and a B\"acklund transformation, which in turn
  leads to a method of constructing soliton solutions. The
  relationship between continuous and discrete equations is discussed
  from two directions: taking the continuum limit of a discrete
  equation and discretizing a continuous equation following the method
  of Hirota.
\end{abstract}

\section{Introduction}
We are all familiar with the integrable (continuous) soliton equations
that have been studied intensively since their resurrection in late
1960's (see e.g. \cite{AC91,DJ89,FaddTakh}). Many interesting and
useful properties are associated with such systems, such as
symmetries, infinite number of conserved quantities, elastic
scattering of solitons, and solvability using various methods
such as the Inverse Scattering Transform and Hirota's bilinear
method. All these nice properties follow from some important
underlying mathematical structure, which has been elaborated in many
studies (e.g. by Mikio Sato and his collaborators in Kyoto, see
e.g. \cite{MJD00}).

With such a beautiful continuous theory of soliton equations, what is
the point of a discrete soliton theory?  One might say that we need to
discretize PDEs in order to do computations with them, or that there
cannot be smooth continuity beyond the Planck scale where quantum
aspects take over. But the reason proposed here is that discrete
soliton equations should be studied because their mathematical
properties of are, if possible, even more beautiful than those of the
continuum equations.

\section{Basic set-up for lattice equations}
When one mentions ``lattice equations'' perhaps the first thing that
comes to mind is the ubiquitous Toda lattice equation given by:
\[
\ddot x_i(t)=e^{-(x_i(t)-x_{i-1}(t))}-e^{-(x_i(t)-x_{i-1}(t))},\quad\forall i.
\]
Here $x_i$ is the position of the particle having the name $i$ and the
equation gives the time evolution $x_i(t)$. Since time is still
continuous we would call this a semi-discrete equation.

\subsection{Equations on Cartesian lattice}
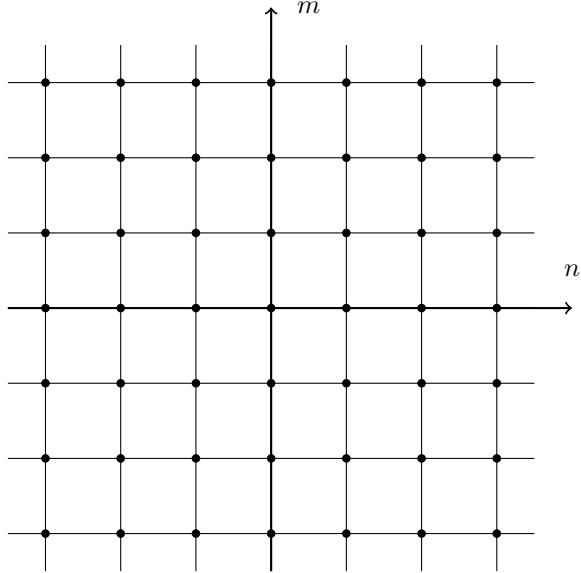
\begin{figure}[h]
\centering
\begin{tikzpicture}[scale=1.0]
\foreach \x in {-3,-2,...,3}{%
      \foreach \y in {-3,-2,...,3}{%
        \node[draw,circle,inner sep=1pt,fill] at (\x,\y) {};}}
\foreach \y in {-3,-2,...,3}%
 \draw[thin] (-3.5,\y) -- (3.5,\y);
\foreach \x in {-3,-2,...,3}%
 \draw[thin] (\x,-3.5) -- (\x,3.5);
\draw[->,thick] (0,-3.5) -- (0,4);
\draw[->,thick] (-3.5,0) -- (4,0);
\node at (0.5,4) {$m$};
\node at (4,0.5) {$n$};
\end{tikzpicture}
\caption{The Cartesian lattice, $u_{n,m}$ are located at lattice
  points.\label{F:1}}
\end{figure}

Here we are considering fully discrete soliton equations and therefore the
continuous $u(x,t)$ will be replaced by $u_{n,m}$, i.e., both the
space and time coordinates are discretized. The most common discrete
two-dimensional space is the Cartesian 2D lattice with dependent
variables located at the vertices of the lattice, see Figure \ref{F:1}.
Other lattices can also be considered, as well as variables not on the
vertices but on the links between the vertices.

\begin{figure}
\centering
\begin{tikzpicture}[scale=2.0]
 \draw[thin] (-0.3,0) -- (1.3,0);
 \draw[thin] (-0.3,1) -- (1.3,1);
 \draw[thin] (0,-0.3) -- (0,1.3);
 \draw[thin] (1,-0.3) -- (1,1.3);
\draw [black] (1,1) circle (1.5pt);
\filldraw [black] (0,0) circle (1.5pt);
\filldraw [black] (1,0) circle (1.5pt);
\filldraw [black] (0,1) circle (1.5pt);
\node at (-0.3,0.16) {$u_{n,m}$};
\node at (1.36,0.16) {$u_{n+1,m}$};
\node at (-0.35,1.16) {$u_{n,m+1}$};
\node at (1.46,1.16) {$u_{n+1,m+1}$};
\end{tikzpicture}
\caption{Equation on a quadrilateral: If values at three corners are
  given, one should be able to compute the value at the fourth
  corner. \label{F:2}}
\end{figure}
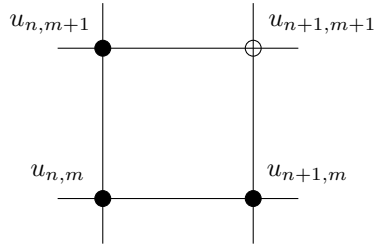

The soliton equations are typically evolution equations and therefore
we must discuss what kind of evolution we can have on the
lattice. The simplest equation is the one relating the
corners of a lattice square or quadrilateral. This involves four
corners and if we give values at three corners we should be able to
compute the value on the fourth corner, see Figure \ref{F:2}.
For this to be possible we must require that the equation is {\em
  affine linear} in all the corner variables. As examples of such
equations we have the lattice potential KdV equation (lpKdV)
\begin{equation}\label{eq:H1}
(u_{n,m}-u_{n+1,m+1})(u_{n,m+1}-u_{n+1,m})=p^2-q^2,\quad\forall n,m,
\end{equation}
and the lattice potential modified KdV equation (lpmKdV)
\begin{equation}\label{eq:H3-0}
p\,(u_{n,m}\,u_{n+1,m}-u_{n,m+1}\,u_{n+1,m+1})=
q\,(u_{n,m}\,u_{n,m+1}-u_{n+1,m}\,u_{n+1,m+1}).
\end{equation}
For these examples we can clearly compute, within each quadrilateral, the
value at any corner once the other three corner values are
given. This is the local situation.

\begin{figure}[h]
\centering
\begin{tikzpicture}[scale=1.0]
\foreach \x in {0,1,2}{%
      \foreach \y in {0,1,2}{%
        \node[draw,circle,inner sep=1.5pt] at (\x,\y) {};}}
\foreach \x in {0,1,2} \node[draw,circle,inner sep=1.5pt,fill] at (\x,-1) {};
\foreach \y in {-1,0,1,2} \node[draw,circle,inner sep=1.5pt,fill] at (-1,\y) {};
\foreach \y in {-2,-1,...,2}%
 \draw[thin] (-2.5,\y) -- (2.5,\y);
\foreach \x in {-2,-1,...,2}%
 \draw[thin] (\x,-2.5) -- (\x,2.5);
\draw[very thick] (-1,-1) -- (-1,2.5);
\draw[very thick] (-1,-1) -- (2.5,-1);
\node at (-0.5,-3) {a)};
\end{tikzpicture}\hspace{2cm}\begin{tikzpicture}[scale=1.0]
\foreach \x in {-2,-1,...,2}
        \node[draw,circle,inner sep=1.5pt,fill] at (\x, - \x) {};
\foreach \x in {-2,-1,...,1}
        \node[draw,circle,inner sep=1.5pt,fill] at (\x, - \x-1) {};

\foreach \z in {-1,0,1,2}{%
\foreach \x in {\z,...,2}
        \node[draw,circle,inner sep=1.5pt] at (\x,-\x+\z+2) {};}
\foreach \x in {-2,-1,...,1}
       \draw[very thick] (\x,-\x) --  (\x,-\x-1); 
\foreach \x in {-2,-1,...,1}
       \draw[very thick] (\x,-\x-1) --  (\x+1,-\x-1); 

\draw[very thick] (-2,2) -- (-2.5,2);
\draw[very thick] (2,-2) -- (2,-2.5);

\foreach \y in {-2,-1,...,2}%
 \draw[thin] (-2.5,\y) -- (2.5,\y);
\foreach \x in {-2,-1,...,2}%
 \draw[thin] (\x,-2.5) -- (\x,2.5);

\node at (-0.5,-3) {b)};
\end{tikzpicture}
\caption{a): The Cartesian lattice with corner initial values (black
  disks) given, one can then compute the values at open circles in the
  upper right quadrant. b) The same with staircase initial values.\label{F:3}}
\end{figure}
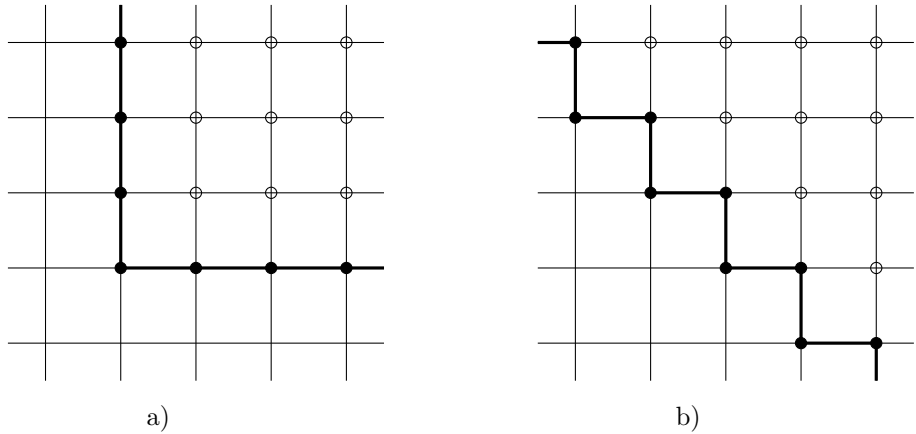

For a global picture it is necessary to define initial values on
some curve so that one can proceed to compute values ``forward''. One
possibility is to give the values on a corner, another is to use
staircase initial values, see Figure \ref{F:3}. In the figure the
evolution is to the upper-right direction, but there are corresponding
initial settings for other directions. Also the staircase can have
occasional longer or higher stairs as long as we have uniquely defined
evolution.

\subsection{Discrete structure within continuous soliton equations}
The examples above \eqref{eq:H1} and \eqref{eq:H3-0} did not come from
thin air.  If we use \eqref{eq:H1} with $n=0,m=0$ and solve for
$u_{1,1}$ we get
\begin{equation}\label{eq:H1-WE}
u_{1,1}=u_{0,0}+\frac{p^2-q^2}{u_{1,0}-u_{0,1}}.
\end{equation}
This may look familiar. Indeed, in 1973 Wahlquist and Estabrook
discussed \cite{WahlEst73} \BA transformation (BT) for KdV solitons
and found (translating notation to the present case) that if $u_{0,0}$
is a ``seed'' solution and $u_{1,0}$ is obtained from it by a BT with
parameter $p$, and similarly $u_{0,1}$ with parameter $q$, then there
is a {\em superposition principle}: If one applies BT with $q$ on
$u_{1,0}$ or with $p$ on $u_{0,1}$ then the results can be the same
(i.e., the BTs commute) and the unique result is given {\em
  algebraically} according to formula \eqref{eq:H1-WE}.

Similar results were derived even before, within the theory of
surfaces. In his studies \BA derived \cite{Back1883} the equation
\[
\theta_{uv}=2\sin(\theta/2),
\] 
that is now called the sine-Gordon equation and subsequently Bianchi
derived \cite{Bianchi1892} the permutability theorem for the BTs
(c.f. \eqref{eq:H1-WE}):
\[
\theta_{12}=\theta+4\,\text{arctan} \left[\frac{\beta_2+\beta_1}{\beta_2-\beta_1}
  \tan\left(\frac{\theta_2-\theta_1}4\right)\right].
\]
If we take $\tan$ on both sides and write the result in terms of
$u:=\exp(i\theta/2)$ we get
\[
\beta_1(u\,u_1-u_2\,u_{12})= \beta_2(u\,u_2-u_1\,u_{12}).
\]
This can again be elevated to an abstract equation on the $\mathbb
Z^2$ lattice, namely to lpmKdV as given in \eqref{eq:H3-0}. Here we
interpret subscripts as giving directions of steps in the lattice.

One can study various properties of abstract lattice equations, but if
they have a connection to continuous soliton equations as noted above,
some of the results may have concrete applications for them.

\section{Symmetries and hierarchies}
\subsection{In the continuum}
One of the essential concepts of integrable soliton equations is that
they do not appear isolated but in hierarchies. For example for the
KdV equation we have the hierarchy of equations
\bse\label{eq:kdvhier}
\begin{eqnarray}
u_{t_1}&=&\partial_xu\\
u_{t_3}&=& \tfrac{1}{ 4}\partial_x\lbrack u_{xx}+3\, u^{2}\rbrack
 \label{eq:potkdv53}  \\
u_{t_5}&=& \tfrac{1}{ 16}\partial_x\lbrack u_{4x}+10\, u_{xx}\, u+5\,
 u_{x}^{2}+10\, u^{3}\rbrack \label{eq:potkdv5} \\
&\vdots&\nn
\end{eqnarray}
\ese
Thus in the KdV case we have one space variable $x$ and multiple time
variables $t_j$, and the flows corresponding to the different times
commute. Furthermore, if we assign weight 2 for $u$, weight 1 for
$\partial_x$ and weight $j$ for $\partial_{t_j}$ then all equations
are weight homogeneous. There are elegant explanations on why the
equations fit together so nicely, e.g. by the Sato theory \cite{MJD00}.

\subsection{Discrete multidimensionality}
For the present discrete case we would also like to have hierarchical
and multidimensional structure. To begin with, our \eqref{eq:H1} is
fully symmetric between the $n$ and $m$ coordinates of the $\mathbb
Z^2$ lattice, and therefore as we introduce higher dimensionality we
would like to keep this symmetry. Thus we introduce a third
dimension and the corresponding lattice index $k$ by $u_{n,m} \to
u_{n,m,k}$ and rewrite \eqref{eq:H1} as
\begin{equation}\label{eq:H1-3nm}
(u_{n,m,k}-u_{n+1,m+1,k})(u_{n,m+1,k}-u_{n+1,m,k})=p^2-q^2,\quad\forall n,m,k.
\end{equation}
This means that we have the same equation on all planes labeled by
$k$. When we look at the situation from this point of view it is
natural to propose \cite{NRGO01} that we should equally well have
equations in which $m$ labels the plane while $n,k$ label the corners
of the quadrilateral:
\begin{equation}\label{eq:H1-3nk}
(u_{n,m,k}-u_{n+1,m,k+1})(u_{n,m,k+1}-u_{n+1,m,k})=p^2-r^2,\quad\forall n,m,k.
\end{equation}
Here we have also replaced $q$ with $r$, which is the lattice constant
for the $k$ direction. Finally we can have a similar equation in the
$m,k$ plane
\begin{equation}\label{eq:H1-3mk}
(u_{n,m,k}-u_{n,m+1,k+1})(u_{n,m,k+1}-u_{n,m+1,k})=q^2-r^2,\quad\forall n,m,k.
\end{equation}

As the subscript notation starts to get lengthy it is common in the
literature to introduce various kinds of shorthand notations. We
sometimes use the notation in which shift in the $n$-direction is
indicated by a tilde, in the $m$-direction by a hat and in the
$k$-direction by a bar:
\begin{align*}
&u_{n,m,k}=u,\quad u_{n+1,m,k}=\t u,\quad u_{n,m+1,k}=\h u,\quad
u_{n,m,k+1}=\b u,\\
& u_{n+1,m+1,k}=\xht u,\quad u_{n+1,m,k+1}=\xbt u,\quad u_{n,m+1,k+1}=\xbh u,\quad
 u_{n+1,m+1,k+1}=\xbht u.
\end{align*}
Then our equations on the three planes read
\bse\label{eq:H1set}\begin{eqnarray}
Q_{12}(u,\t u,\h u,\xht u;p,q)&:=&(u-\xht u)(\h u-\t u)-p^2+q^2=0,\\
Q_{23}(u,\h u,\b u,\xbh u;q,r)&:=&(u-\xbh u)(\b u-\h u)-q^2+r^2=0,\\
Q_{31}(u,\b u,\t u,\xbt u;r,p)&:=&(u-\xbt u)(\t u-\b u)-r^2+p^2=0,
\end{eqnarray}\ese
when written using cyclic changes: tilde $\to$ hat $\to$ bar, $p\to q\to r $.

\subsection{Commuting discrete flows}
\begin{figure}
\centering
\begin{tikzpicture}[scale=1.70,tdplot_main_coords, cube/.style={thick,black,inner sep=0pt}]
\draw[cube] (0,2,0) -- (2,2,0) -- (2,0,0);
\draw[cube] (0,0,2) -- (0,2,2) -- (2,2,2) -- (2,0,2) -- cycle;
        \draw[dashed] (0,0,0) -- (0,0,2);
        \draw[dashed] (0,0,0) -- (0,2,0); 
        \draw[dashed] (0,0,0) -- (2,0,0);
       \draw[cube] (0,2,0) -- (0,2,2);
        \draw[cube] (2,0,0) -- (2,0,2);
        \draw[cube] (2,2,0) -- (2,2,2);
\node at (0,0,-0.2) {$u$};
\node at (2,0,-0.2) {$\t u$};
\node at (2,2.1,-0.2) {$\xht u$};
\node at (0,2,-0.2) {$\h u$};
\node at (0,0.2,2.2) {$\b u$};
\node at (2,0,2.3) {$\xbt u$};
\node at (2,2,2.3) {$\xbht u$};
\node at (0,2,2.3) {$\xbh u$};
\draw[thick,->] (2,0,0) -- (3,0,0);
\draw[thick,->] (0,2,0) -- (0,2.7,0);
\draw[thick,->] (0,0,2) -- (0,0,2.7);
\node at (0,0.4,2.7) {$r,\, k$};
\node at (3,-0.4,0) {$p,\, n$};
\node at (0,2.7,0.2) {$q,\, m$};
\draw[black,fill] (0,0,0) circle [radius=0.05cm];
\draw[black,fill] (0,0,2) circle [radius=0.05cm];
\draw[black,fill] (0,2,0) circle [radius=0.05cm];
\draw[black,fill] (2,0,0) circle [radius=0.05cm];

\draw[gray!50,fill] (2,2,0) circle [radius=0.05cm];
\draw[black] (2,2,0) circle [radius=0.051cm];
\draw[gray!50,fill] (2,0,2) circle [radius=0.05cm];
\draw[black] (2,0,2) circle [radius=0.051cm];
\draw[gray!50,fill] (0,2,2) circle [radius=0.05cm];
\draw[black] (0,2,2) circle [radius=0.051cm];

\draw[gray!10,fill] (2,2,2) circle [radius=0.07cm];
\draw[black] (2,2,2) circle [radius=0.071cm];

\end{tikzpicture}
\caption{The consistency cube. Evolution can start on each plane from
  the corner with values at the black disks given. The values at gray
  circles can then be computed uniquely, but the value at the open
  circle may be ambiguous as it can be computed in three different
  ways.\label{F:CACcube}}
\end{figure}
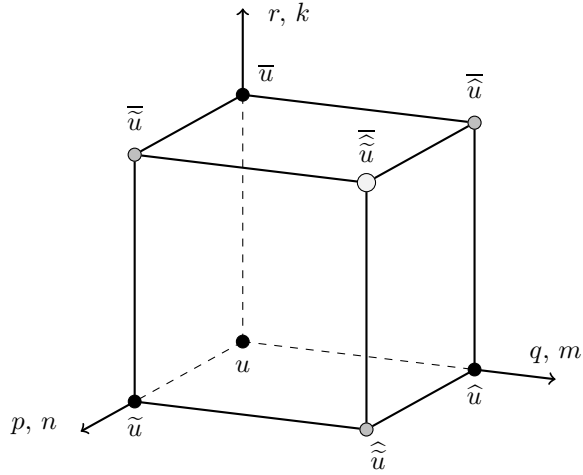

In the continuum case we know that we cannot introduce arbitrary flows
in the different time directions because they would not be compatible, i.e.,
they would not commute. We have already discussed evolution on the
lattice (see Figure \ref{F:3}) and when we assign equations on
different planes, the evolution they generate must also satisfy some
compatibility conditions. Let us look at this locally. Assuming a
common corner $(n,m,k)$ in the $\mathbb Z^3$ lattice we should have
a situation as in Figure \ref{F:2} in each of the three planes
intersecting in that corner. If we keep just the elementary plaquettes
we get Figure \ref{F:CACcube}.

\tdplotsetmaincoords{75}{115}

In terms of equations the situation is as follows: At the various sides
of the cube we have the corresponding equations:
\bse\label{eq:conseqs}\begin{align}
\text{bottom:}\quad & Q_{12}(u,\t u,\h u,\xht u;p,q)=0.&
\text{top:}\quad &Q_{12}(\b u,\xbt u,\xbh u,\xbht u;p,q)=0,&\\ 
\text{back:}\quad & Q_{23}(u,\h u,\b u,\xbh u;q,r)=0,&
\text{front:} \quad &Q_{23}(\t u,\xht u,\xbt u,\xbht u;q,r)=0,&\\
\text{left:}\quad & Q_{31}(u,\b u,\t u,\xbt u;r,p)=0,&
\text{right:} \quad &Q_{31}(\h u,\xbh u,\xht u,\xbht u;r,p)=0.&
\end{align}
\ese Here we get from the LHS to the RHS by applying on the dependent
variables a shift in the direction not yet appearing on the LHS while
keeping the equation itself unchanged.  We would use this for any
candidate equations which are uniform on parallel planes, for lpKdV we
have \eqref{eq:H1set}.

From a corner we can start evolution and for the configuration of
Figure \ref{F:CACcube} with $u,\t u,\h u,\b u$ as initial values we
can compute using LHS equations the values of $\,\xht u,\, \xbt u,\,
\xbh u$.  After this we can compute $\xbht u$ from each of the three
RHS equations and the result should be the same. In the language of
the commuting flows we have three different order of flows
\begin{itemize}
\item first, independently, (LHS $Q_{12}$ to get $\xht u$, and LHS $Q_{23}$ to
  get $\xbh u$ ), after that RHS $Q_{31}$ to get $\xbht u$.
\item first, independently, (LHS $Q_{23}$ to get $\xbh u$, and LHS $Q_{31}$ to
  get $\xbt u$ ), after that RHS $Q_{12}$ to get $\xbht u$.
\item first, independently, (LHS $Q_{31}$ to get $\xbt u$, and LHS $Q_{12}$ to
  get $\xht u$ ), after that RHS $Q_{23}$ to get $\xbht u$.
\end{itemize}
Thus we have three flows, which can be arranged in 6 different orders, but
since the order in the first pair does not matter we find the
condition that the three possibilities listed above should give the
same result, i.e., two consistency conditions. This is also called
``Consistency-Around-a-Cube'' (CAC) or Multidimensional consistency
(MDC). When this is applied to equations \eqref{eq:H1set}
we find that in each case
\[
\xbht u=-\,\frac{
\t u\h u\,(p^2-q^2)+\h u\b u\,(q^2-r^2)+\b u\t u\,(r^2-p^2)}{
\t u\,(q^2-r^2)+\h u\,(r^2-p^2)+\b u\,(p^2-q^2)}.
\]
This was already derived by Wahlquist and Estabrook in the context of
BT \cite{WahlEst73}.

In the general case the conditions following from CAC are fairly
complicated, but under suitable additional assumptions one can obtain
a classification of equations, the most interesting being the ``ABS
list'' \cite{ABS03}, which contains the above mentioned lpKdV as
``H1'' and lpmKdV as ``H3($\delta=0$)''.

\section{Lax pairs}
\subsection{Constructing the Lax pair from CAC}
In the discrete case we can use the equations on the
consistency cube to generate a Lax pair by taking the bar-variables as
the auxiliary linear variables.

Let us take the back and left equations of \eqref{eq:conseqs} and
solve for $\xbh u$ and $\xbt u$. In the case of lpKdV we get
\bse\label{eq:Lax1}\begin{eqnarray}
\xbh u&=&\frac{u(\b u-\h u)-q^2+r^2}{\b u-\h u},\\
\xbt u&=&\frac{u(\b u-\t u)-p^2+r^2}{\b u-\t u},
\end{eqnarray}
\ese
Now introducing
\[
\b u=\frac{f}{g}
\]
we can write \eqref{eq:Lax1} as
\bse\label{eq:Lax1m}\begin{eqnarray}
\frac{\h f}{\h g}&=&\frac{u\,f-(u\h u+q^2-r^2)\,g}{f-\h u\, g},\\
\frac{\t f}{\t g}&=&\frac{u\,f-(u\t u+p^2-r^2)\,g}{f-\t u\, g}.
\end{eqnarray}
\ese 
This can be written as a matrix equation:
\[
\h \Phi={\cal M}\Phi,\quad
\t \Phi={\cal L}\Phi,\quad \Phi:=\begin{pmatrix} f \\ g \end{pmatrix},
\]
where
\begin{equation}\label{eq:Lax2}
{\cal M}:=\mu(u,\h u)\begin{pmatrix} u & -(u \h u+q^2-r^2) \\
1 & -\h u \end{pmatrix},
\quad
{\cal L}:=\lambda(u,\t u)\begin{pmatrix} u & -(u\t u+p^2-r^2) \\
1 & -\t u \end{pmatrix}.
\end{equation}
(Here $\mu(u,\h u)$ and $\lambda(u,\t u)$ are separation
factors, one way to fix them is to require that
$\det{\cal M}=1,$ $\det{\cal L}=1$.) The compatibility condition arises 
from 
\[
(\,\h\Phi\,)\t{\phantom\Phi}=
(\,\t\Phi\,)\h{\phantom\Phi}
\]
which implies 
\begin{equation}\label{eq:comm}
\t{\cal M}\,{\cal L}=\h {\cal L}\,{\cal M}.
\end{equation}
Applying the matrices given in \eqref{eq:Lax2} (with
$\mu=\lambda=1$) to this equation yields
\[
\t{\cal M}\,{\cal L}-\h {\cal L}\,{\cal M}=
\left[(u-\xht u)(\h u-\t u)-p^2+q^2\right]
\begin{pmatrix}-1 & \h u+\t u \\ 0 & 1 \end{pmatrix},
\]
and thus the Lax pair does generate the equation. However, it should
be noted that there can also be ``fake Lax pairs'', that is, even if
an equation has the CAC property its Lax pair as constructed above
might not generate the equation (for example if equation
\eqref{eq:comm} is satisfied automatically).

\subsection{\BA transformation for constructing soliton solutions} 
The Lax pair and the \BA transformation are different ways
of interpreting the six equations \eqref{eq:conseqs}: For a Lax pair we
used the back and left equations and wrote them in matrix form. For BT
we assume that $u_{n,m,1}$ solves the top equation and then we use
the back and left equations to construct a solution to the bottom
equation (which has the same form as the top equation). Since we
have the extra lattice parameter $r$ at our disposal, the solution to
the bottom equation should be more general.

The starting point in this construction is a seed solution of the
top equation. Usually this is just the null solution, but now we
observe that $u_{n,m,k}\equiv 0$ is not a solution of
\eqref{eq:H1-3nm} and the first problem is to find a suitable seed
solution. One finds easily that
\[
u_{n,m,k}=\pm p\,n\pm q\,m+c\,k
\]
is a simple linear solution. With this in mind let us change dependent
variables by
\begin{equation}\label{eq:u2v}
u_{n,m,k}=v_{n,m,k}-pn-qm-rk
\end{equation}
after which the bottom, back and left equations can be written,
respectively, as \bse\begin{eqnarray} 
(v_{n+1,m,k}-v_{n,m+1,k} -p+q)
(v_{n+1,m+1,k}-v_{n,m,k}-p-q)&=&(p^2-q^2),\phantom{mmmm}\label{eq:vbot}\\ 
(v_{n,m+1,k}-v_{n,m,k+1} -q+r)
(v_{n,m+1,k+1}-v_{n,m,k}-q-r)&=&(q^2-r^2),\\ 
(v_{n,m,k+1}-v_{n+1,m,k} -r+p)
(v_{n+1,m,k+1}-v_{n,m,k}-r-p)&=&(r^2-p^2).
\end{eqnarray}\ese
We now use these equations for $k=0$, take $v_{n,m,1}=0,\,\forall n,m$,
which solves the top equation, and solve for
$\nu_{n,m}:=v_{n,m,0}$. We find
\begin{eqnarray}
\nu_{n,m+1}&=&\frac{(q-r)\,\nu_{n,m}}{\nu_{n,m}+q+r},\\
\nu_{n+1,m}&=&\frac{(p-r)\,\nu_{n,m}}{\nu_{n,m}+p+r}.
\end{eqnarray}
Again we would like to use matrix notation to write these results, and for
that purpose we introduce
\[
\psi_{n,m}=\begin{pmatrix}a_{n,m} \\ b_{n,m} \end{pmatrix},\quad
M:=\mu\begin{pmatrix}q-r & 0 \\ 1 & q+r\end{pmatrix},\quad
L:=\lambda\begin{pmatrix}p-r & 0 \\ 1 & p+r\end{pmatrix},
\]
so that the equations to solve are
\[
\psi_{n,m+1}=M\,\psi_{n,m},\quad\psi_{n+1,m}=L\,\psi_{n,m}.
\]
Since $M,L$ are commuting constant matrices and
\[
M^m=\begin{pmatrix}F^m &
0\\(1-F^m)/(2r) & 1\end{pmatrix},\quad
L^n=\begin{pmatrix}G^n &
0\\(1-G^n)/(2r) & 1\end{pmatrix}
\]
where $F:=(q-r)/(q+r)$, $G:=(p-r)/(p+r)$,
we find
\[
\psi_{n,m}=M^m\,L^n\,\psi_{0,0}.
\]
Putting everything together yields the result
\begin{equation}\label{eq:rho}
\nu_{n,m}=\frac{a_{n,m}}{b_{n,m}}=2r\frac{\rho_{n,m}}{1-\rho_{n,m}},
\quad\text{where}\quad
\rho_{n,m}=\left(\frac{q-r}{q+r}\right)^m\left(\frac{p-r}{p+r}\right)^n
\frac{v_{0,0}}{2r+v_{0,0}}.
\end{equation}

\section{Continuum limits}
\begin{figure}
\centering
\begin{tikzpicture}[scale=1.5]
\coordinate (q) at (1,-0.4); 
 \draw[thin] (0,1) -- (1,1) -- (1,0) -- (0,0) -- (0,1);
\filldraw [black] (1,1) circle (1.5pt);
\filldraw [black] (0,0) circle (1.5pt);
\filldraw [black] (1,0) circle (1.5pt);
\filldraw [black] (0,1) circle (1.5pt);
\node at (-0.2,-0.2) {$u_{n,m}$};
\node at (1.2,-0.2) {$u_{n+1,m}$};
\node at (-0.2,1.2) {$u_{n,m+1}$};
\node at (1.2,1.2) {$u_{n+1,m+1}$};
\end{tikzpicture}\hspace{1.5cm}\begin{tikzpicture}[scale=1.5,baseline=(q)]
\draw[thin] (0,0) -- (1,0) -- (1,0.5) -- (0,0.5) -- (0,0);
\filldraw [black] (1,0.5) circle (1.5pt);
\filldraw [black] (0,0) circle (1.5pt);
\filldraw [black] (1,0) circle (1.5pt);
\filldraw [black] (0,0.5) circle (1.5pt);
\end{tikzpicture}\hspace{1.5cm}
\begin{tikzpicture}[scale=1.5,baseline=(q)]
 \draw[thin] (0,0) -- (1,0);
\phantom{  \draw[thin] (0,-0.3) -- (0,0.3);}
\phantom{   \draw[thin] (1,-0.3) -- (1,0.3);}
\filldraw [black] (0,0) circle (1.5pt);
\filldraw [black] (1,0) circle (1.5pt);
\node at (-0.3,0.3) {$ u_{n}(\tau),\,\dot u_{n}(\tau)$};
\node at (1.3,-0.3) {$ u_{n+1}(\tau),\,\dot u_{n+1}(\tau)$};
\end{tikzpicture}\vspace{1cm}\hspace{1cm}
\begin{tikzpicture}[scale=1.2]
\coordinate (p) at (1,-1.2);
 \draw[thin] (0,0) -- (1,1) -- (2,0) -- (1,-1) -- (0,0);
\filldraw [black] (1,1) circle (1.8pt);
\filldraw [black] (2,0) circle (1.8pt);
\filldraw [black] (1,-1) circle (1.8pt);
\filldraw [black] (0,0) circle (1.8pt);
\node at (-0.5,0.2) {$u_{n'-1,m'}$};
\node at (1.8,1) {$u_{n',m'+1}$};
\node at (1.8,-1) {$u_{n',m'-1}$};
\node at (2.7,-0.2) {$u_{n'+1,m'}$};
\end{tikzpicture}\hspace{0.5cm}\begin{tikzpicture}[scale=1.2,baseline=(p)]
 \draw[thin] (0,0) -- (1,0.4) -- (2,0) -- (1,-0.4) -- (0,0);
\filldraw [black] (1,0.4) circle (1.8pt);
\filldraw [black] (2,0) circle (1.8pt);
\filldraw [black] (1,-0.4) circle (1.8pt);
\filldraw [black] (0,0) circle (1.8pt);
\end{tikzpicture}\hspace{0.5cm}\begin{tikzpicture}[scale=1.2,baseline=(p)]
 \draw[thin] (0,0) -- (2,0);
\filldraw [black] (1,0) circle (1.8pt);
\filldraw [black] (2,0) circle (1.8pt);
\filldraw [black] (0,0) circle (1.8pt);
\node at (-0.3,0.3) {$u_{n'-1}(t)$};
\node at (1.1,-0.3) {$u_{n'}(t),\,u'_{n'}(t)$};
\node at (2.4,0.3) {$u_{n'+1}(t)$};
\end{tikzpicture}
\caption{There are two ways to squeeze the quadrilateral to obtain
  continuum limits. $\dot u$ and $u'$ are the corresponding
  derivatives. \label{F:clim}}
\end{figure}
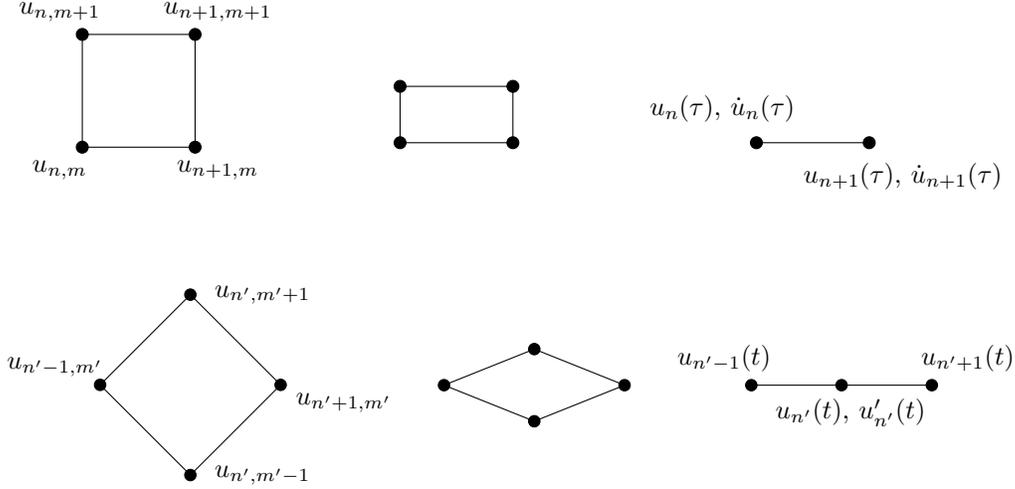

When we compare discrete and continuous spaces we will match the
origins and then for a generic point we have $(x,y)=(\epsilon n,\delta
m)$, where $\epsilon$ and $\delta$ measure the lattice distances. For
a quad equation there are two ways to take a continuum limit as
illustrated in Figure \ref{F:clim}: In the top figure (straight limit)
the square is squeezed in the $m$-direction, in the bottom figure
(skew limit) it is first rotated $45^\circ$ and then squeezed.

\subsection{Skew limit}
Here we will only consider the skew limit. For that purpose we
rotate the coordinates by $(n,m)\to(n+m-1,m-n)$, furthermore let us
denote $n+m=n',\,m-n=m'$ and then equation \eqref{eq:vbot} reads
\begin{equation}\label{eq:vpeq}
(v_{n',m'-1}-v_{n',m'+1}-p+q)(v_{n'+1,m}-v_{n'-1,m'}-p-q)=p^2-q^2.
\end{equation} 
Since we take the limit in the $m'$ direction we set
\begin{equation}\label{eq:vclim}
v_{n'+\nu,m'+\mu}=V_{n'+\mu}(t+\delta \mu),
\end{equation}
where $\delta$ is the lattice distance in the $m'$ direction.  Thus we
will take 
\begin{equation}\label{eq:sqlim}
m'\to\infty,\quad \delta\to 0,\quad\text{while}\quad m'\delta=t
\quad\text{stays fixed}.
\end{equation}

We still have the question of how $\delta$ and $p,q$ are related. Some
help can be obtained from the form of $\rho$ in \eqref{eq:rho}. We 
know that the soliton solutions are constructed using exponential
functions and $\rho$ can be interpreted as a discrete form of the
exponential, due to the well known limit formula
\begin{equation}\label{eq:dexp}
\lim_{n\to\infty}\left(1+\frac{x}n\right)^n= e^x.
\end{equation}
In our case we have to consider the combination with $m-n$ power, i.e.
\[
\left(\frac{q-r}{q+r}\cdot\frac{p+r}{p-r}\right)^{(m-n)}=
\left(1+\frac{2r(q-p)}{(p-r)(q+r)}\right)^{m'}
\]
Since $r$ is a soliton parameter it will stay finite and nonzero and
therefore we take $q-p=\delta\to 0$. Substituting $q=p+\delta$ and using
\eqref{eq:sqlim} we get 
\[ 
\left(1+\frac{2r\delta}{(p-r)(p+\delta+r)}\right)^{m'}\hspace{-0.2cm}=
\left(1+\frac{t}{m'}\frac{2r}{(p-r)(p+t/m'+r)}\right)^{m'}\hspace{-0.2cm}\to\,
\exp\left(\frac{2rt}{p^2-r^2}\right).
\]
Thus the limit works and produces a reasonable plane wave factor. We
then proceed to insert $q=p+\delta$ and \eqref{eq:vclim} into
\eqref{eq:vpeq} and expand in $\delta$. This yields
\begin{equation}
\partial_t V_n(t)=1-\frac{2p}{V_{n-1}(t)-V_{n+1}(t)+2p}
\end{equation}
where we have dropped the primes in $n$. This equation is therefore
the skew semi-discrete limit of the (translated) lpKdV equation given in
\eqref{eq:vbot}. It is a bona-fide integrable equation, having a Lax
pair etc.

We can next take a continuum limit in the remaining $n$ variable. For
this purpose we take $p=1/\epsilon$ and write
\[
V_{n+\nu}(t)=U(t,\xi+\nu\epsilon)
\]
and expand in epsilon. The result is
\begin{eqnarray*}
\partial_t U &=&\epsilon^2 U_x\\
&&+\epsilon^4\tfrac16[U_{xxx}+6U_x^2]\\
&&+\text{ h.o.}
\end{eqnarray*}
That does not work, the leading term in this limit is not
KdV. However, if we change from $t$ to a scaled-translated variable
$\tau$ by
\begin{equation}\label{eq:sqvar}
\partial_t=\epsilon^2\partial_x+\epsilon^4\partial_\tau
\end{equation}
then the leading term $\epsilon^4$ yields
\begin{equation}
U_\tau=\tfrac16[U_{xxx}+6U_x^2],
\end{equation}
which is a potential form of the KdV equation (pKdV). The need for some sort
of new ``squeezed'' variables as in \eqref{eq:sqvar}  is obvious: the
starting discrete equation is very symmetric while the continuum
target equation is asymmetric, with $x$ playing a different role in
comparison to the $t_i$.

\subsection{Double limit}
On the basis of the above we could try to take a limit in $n,m$
directions simultaneously, but at the same time we should somehow
introduce suitable scaling. Thus we try
\begin{equation}\label{eq:dbl}
v_{n,m}=V(x+\epsilon(n a_1+m b_1),t+\epsilon^3(n a_3+m b_3)),
\end{equation}
where we have chosen the powers of $\epsilon$ following the expected
relative scaling of $x$ and $t_3$. Inserting this with
$p=\alpha/\epsilon,\,q=\beta/\epsilon$ into \eqref{eq:vbot} and
expanding in $\epsilon$ we find that if we choose
\begin{equation}\label{eq:dbllim}
a_j=\frac{2^j}{j \alpha^j},\quad b_j=\frac{2^j}{j \beta^j},\quad
\alpha^2\neq \beta^2,
\end{equation}
we get, as the leading term, the pKdV equation in the form
\begin{equation}\label{eq:expeq3}
V_t=\tfrac14[V_{xxx}+3V_x^2].
\end{equation}

But there is more: If we look at the next order in $\epsilon$ we find
an $x$ derivative of the above equation, and then at the next order,
after using \eqref{eq:expeq3} to eliminate time derivatives, the expression
\[
V_{xxxxx}+10\, V_{xxx}\,V_x+5\, V_{xx}^2+10 V_x^3,
\]
which appears in the square brackets on the RHS of the fifth order KdV
\eqref{eq:potkdv5}. Thus it seems that the discrete equation contains
inside it the whole hierarchy of continuum equations! In order to
explore this further, let us use multiple time variables as follows:
\[
v_{n,m}=V\left(x+\epsilon(n a_1+m b_1),t_3+\epsilon^3(n a_3+m b_3),
t_5+\epsilon^5(n a_5+m b_5),
\cdots\right)
\]
When we expand \eqref{eq:vbot} using this multi-time expression with
parameters \eqref{eq:dbllim} and in the results eliminate lower order
times using lower order equations, and change $V_x=u$, we get the
sequence of higher order members of the KdV hierarchy, some of which
were given in \eqref{eq:kdvhier}.

The above observations can be made into precise statements using
more refined mathematics, for example by using the Sato theory. In
that formalism infinite number of time variables are used at the
outset and one can find a simple correspondence between the discrete
and continuum hierarchies. The main observation to take away from this
is that the nicely symmetric and simple looking discrete formalism is
in effect as rich as the corresponding continuum theory. And this
statement holds also for the more general equations such as KP.

\section{Discretizing a continuous equation}
One approach to discrete equations is to take a known continuous
integrable equation and try to construct a discrete version with as
many as possible integrability properties. One important object that
we would like to preserve is the class of solutions, perhaps in the
sense that the discrete solutions approach the continuous ones in a
smooth fashion.

\subsection{A simple 1D example}
Consider the nonlinear ODE (Verhulst's population model)
\begin{equation}\label{eq:Veq}
\dot x(t)=\alpha\, x(t)(1-x(t)).
\end{equation}
We would like to discretize this so that the solutions of the discrete
version follow closely the continuous solution, which can be derived
easily:
\begin{equation}\label{eq:Vcsol}
x(t)=\frac{1}{1+e^{\alpha\, (t-t_0)}}.
\end{equation}

How should this equation be discretized? A naive discretization would
be to replace the derivative by a forward difference:
\begin{equation}\label{eq:Vnaive}
h^{-1}(x(t+h)-x(t))=\alpha\, x(t)(1-x(t)).
\end{equation}
This is the logistic equation which is well known to lead to chaotic
behavior for most values of the parameter $\alpha$, while the solution
\eqref{eq:Vcsol} is always smooth. We need a different discretization.

In order to proceed we note that equation \eqref{eq:Veq} can be
linearized:
\begin{equation}\label{eq:dlinver}
x(t)=\frac1{1+y(t)}\quad \Rightarrow\quad \dot y(t)+\alpha\,y(t)=0.
\end{equation}
For the linear equation the naive discretization works: The solution
to the continuous $y$ equation \eqref{eq:dlinver} is given by
\begin{equation}\label{eq:Vlincont}
y(t)=\exp[-\alpha(t-t_0)]
\end{equation}
while the solution to the discretized version of \eqref{eq:dlinver}
\[
h^{-1}(y(t+h)-y(t))+\alpha\,y(t)=0
\]
is given by 
\begin{equation}\label{eq:Vlindiskr}
y(t+nh)=(1-\alpha\,h)^{n+(t-t_0)/h}.
\end{equation}
This solution approximates the solution \eqref{eq:Vlincont}, due to
the limit formula \eqref{eq:dexp}:
\[
(1-\alpha h)^{(t-t_0)\frac1h}\,\to\, e^{-\alpha(t-t_0)}\quad
\text{as}\quad h\to 0^+.
\]

Let us denote $y(t+nh)=y_n,\,(t-t_0)/h=-n_0$ and reverse the steps
above. We find
\[
y_n:=(1-\alpha\,h)^{n-n_0}\quad\text{solves}\quad y_{n+1}=(1-\alpha\,h) y_n
\]
and since $x=1/(1+y)$,
\bse\begin{equation}\label{eq:Vxdsol}
x_n:=\frac1{1+(1-\alpha\,h)^{n-n_0}},
\end{equation}
solves
\begin{equation}\label{eq:Vxdeq}
x_{n+1}=\frac{x_n}{1-\alpha\,h+\alpha\,h\,x_n}.
\end{equation}
\ese The solution for $x_n$ \eqref{eq:Vxdsol} is a good approximation
to \eqref{eq:Vcsol} but the equation it solves \eqref{eq:Vxdeq} is not
at all like the one obtained by naive discretization
\eqref{eq:Vnaive}.

\subsection{Hirota's method of discretization}
For PDE's the situation is much more complicated. This is the case in
particular because we do not know all solutions, or rather, the
general solution is too complicated to work with. One approach is to
make sure that at least the soliton solutions carry over from
continuous to discrete. For this we follow R. Hirota, who in a series
of papers in 1977 discretized many soliton equations while preserving
their $N$-soliton solutions \cite{Hir77a,Hir77b,Hir77c}. The
culmination of this work was the ``DAGTE'' equation \cite{Hir81} from
which many other soliton equations follow.

\subsubsection{Bilinear form of continuous KdV}
Hirota's method is based on a transformation of the dependent variable
so that in terms of the new dependent variable the soliton solutions are
simply polynomials of exponentials with linear exponents. Instead of the
standard form of the KdV equation
\begin{equation}\label{eq:Hkdv}
u_t+6\,u\,u_x+u_{xxx}=0.
\end{equation}
it is better to introduce the variable $v$ by $u=v_x$, 
and integrate \eqref{eq:Hkdv} into the pKdV equation
\begin{equation}\label{eq:Hkdvpot}
v_t+3v_x^2+v_{xxx}=0.
\end{equation} 
The new dependent variable $f$ is defined by
\begin{equation}\label{eq:nl2bil}
v=2\partial_x\log(f),\quad\text{or}\quad u=2\partial_x^2\log(f),
\end{equation}
and when this is used in \eqref{eq:Hkdvpot} we get a fourth order
equation
\begin{equation}\label{eq:Hbilkdv}
D_x(D_t+D_x^3)\,f\cdot f=0,
\end{equation}
which is written in terms of Hirota's bilinear derivatives, defined by
\[
D_x^n\,D_t^m\,f\cdot g=\left.(\partial_x-\partial_{x'})^n
(\partial_t-\partial_{t'})^m\,f(x,t)\, g(x',t')\right|_{x'=x,\,y'=y}.
\]

\subsubsection{Discretizing KdV}
In order to continue we need a discrete version of the bilinear
derivative. For usual derivatives we have
\[
e^{ax}f(x)=f(x+a)
\]
and therefore we have, for example,
\begin{eqnarray*}
e^{a D_x}\,f\cdot g&=&f(x+a)\,g(x-a),\\
\sinh(a D_x)f\cdot g&=&\tfrac12[f(x+a)g(x-a)-f(x-a)g(x+a)].
\end{eqnarray*}
Since 
\[
\sinh(a D_x)=a D_x+ \text{ higher order terms in $a D_x$}
\]
it seems reasonable to use discretization rules like $D \to \sinh(a
D)$. The precise replacement to \eqref{eq:Hbilkdv} proposed by Hirota
was (ref \cite{Hir77a}, equation (2.3))
\begin{equation}\label{eq:doHkdv}
\sinh[\tfrac14(D_x+\delta D_t)]\left\{2\delta^{-1}\sinh[\tfrac12\delta D_t]
+2\sinh[\tfrac12 D_x]\right\}\,f(x,t)\cdot f(x,t)=0,
\end{equation}
which can also be written as
\begin{align*}
&\left\{\cosh[\tfrac14\delta D_t+\tfrac34 D_x]
+\delta^{-1}\cosh[\tfrac34\delta D_t+\tfrac14 D_x]\right.\\
&\hspace{2.5cm}-\left.(1+\delta^{-1})\cosh[\tfrac14\delta D_t-\tfrac14 D_x]\right\}
\,f(x,t)\cdot f(x,t)=0.
\end{align*}
In order to write this as shifts we note that
\[
\cosh(\alpha D_x+\beta D_t)\,f(x,t)\cdot f(x,t)=f(x+\alpha,t+\beta)
f(x-\alpha,t-\beta)
\]
and if we convert shifts to discrete subscript notation
\[
f(x+\tfrac14\nu,t+\delta\tfrac14\mu)=f_{n+\frac14\nu,m+\frac14\mu},
\]
we can write \eqref{eq:doHkdv} as
\begin{align}\label{eq:dHkdv}
&f_{n+\frac34,m+\frac14}f_{n-\frac34,m-\frac14}+
\delta^{-1}f_{n+\frac14,m+\frac34}f_{n-\frac14,m-\frac34}
\nn\\&\hspace{4.5cm}
-(1+\delta^{-1})f_{n-\frac14,m+\frac14}f_{n+\frac14,m-\frac14}=0.
\end{align}
This does not sit at the points of the $\mathbb Z^2$ lattice but if we
make a $45^\circ$ rotation and a shift according to
\[
(n+\nu,m+\mu)\mapsto(n+m+\nu+\mu,n-m+\nu-\mu+1)=(n'+\nu+\mu,m'+\nu-\mu)
\]
we get
\begin{equation}\label{eq:dHkdvrot}
f_{n'+1,m'+1}f_{n'-1,m'}+\delta^{-1}f_{n'+1,m'}f_{n'-1,m'+1}
-(1+\delta^{-1})f_{n',m'}f_{n',m'+1}=0.
\end{equation}
The dependent variables are now on the points of the $\mathbb Z^2$
lattice, but the equation connects points on two quadrilaterals.
(This
is typical for Hirota bilinear equations, in fact the only one-component
equation that can exist on a single quadrilateral is trivial.)

Equations \eqref{eq:dHkdv} and \eqref{eq:dHkdvrot} have the main
properties essential  in Hirota's approach to constructing soliton
solutions: a) $f_{n,m}\equiv 1$ is a solution, and b) in each product
the sum of indices is the same. This last property implies gauge
invariance: if $f_{n,m}$ is a solution, so is $f_{n,m}':=A^n
B^m\,f_{n,m}$ for any constants $A,B$.

\subsubsection{Soliton solutions}
In Hirota's approach soliton solutions are constructed perturbatively:
\paragraph{Background solution:}
The bilinear form \eqref{eq:dHkdvrot} obviously has $f_{n,m}\equiv 1$ as
the vacuum or background solution.

\paragraph{One-soliton solution:}
The ansatz for the one-soliton solution of \eqref{eq:dHkdvrot} is
\begin{equation}\label{eq:1ss}
f_{n,m}=1+c A(p,k_1)^n B(q,k_1)^m,
\end{equation}
where $k_1$ is the parameter of the soliton. We have also noted possible
dependence on lattice parameters: the {\em plane wave factor} $A$ may
depend on $p$ because it is associated with the $n$ direction,
similarly $B$ may depend on $q$. The constant $c$ is constant only in
$n,m$ but may depend on $p,q,k_1$.  This ansatz leads to the {\em
  dispersion relation}
\[
A(p,k_1)-B(q,k_1)=\delta\,[1-A(p,k_1)B(q,k_1)],
\]
and evidently $\delta$ should also depend on $p,q$. This relation is
resolved by
\begin{equation}\label{eq:dkdvplane}
A(p,k_1)=\frac{p-k_1}{p+k_1},\quad
B(q,k_1)=\frac{q-k_1}{q+k_1},\quad
\delta(p,q)=\frac{p-q}{p+q}.
\end{equation}
Note the beautiful symmetry which even encompasses the parameter $\delta$.

\paragraph{Two-soliton solution:}
Following Hirota's perturbative approach the 2SS ansatz is
\begin{eqnarray}\label{eq:2ss}
f_{n,m}&=&1+c_1 A(p,k_1)^n B(q,k_1)^m+c_2 A(p,k_2)^n B(q,k_2)^m\nn\\
&&+{\cal A}(k_1,k_2)c_1c_2 A(p,k_1)^n B(q,k_1)^m A(p,k_2)^n B(q,k_2)^m.
\end{eqnarray}
This form is dictated by the condition that when solitons are far
apart they look like 1SS. There is a new parameter ${\cal A}(k_1,k_2)$
called the {\em phase factor}.  When this ansatz is substituted into
\eqref{eq:dHkdvrot} with \eqref{eq:dkdvplane}, we find that it is a
solution, provided that the phase factor is given by
\[
{\cal A}(k_1,k_2)=\frac{(k_1-k_2)^2}{(k_1+k_2)^2}.
\]
This is exactly same as for continuous KdV equation.

\paragraph{$N$-soliton solution in determinant form:} We could follow 
this perturbative route and construct an ansatz for 3SS, with only
$k_3$ as a new parameter, and verify that it is a solution. But we
can do better and construct a general determinant formula for the
$N$-soliton solution.  For that purpose let us define
\[
\psi_{n,m}(j,l):=\rho^+_j\,(p+k_j)^{n}\,(q+k_j)^{m}\,k_j^{l}
+\rho^-_j\,(p-k_j)^{n}\,(q-k_j)^{m}\,(-k_j)^{l}.
\]
It is easy to verify that
$\psi_{n,m}(1,0)$ is gauge equivalent to $f_{n,m}$ of \eqref{eq:1ss} for
$c=\rho^-_1/\rho^+_1$. With $\psi$ given we write the 2SS as
\[
f_{n,m}^{[2ss]}=\left|\begin{matrix} \psi_{n,m}(1,0) & \psi_{n,m}(1,1)\\
 \psi_{n,m}(2,0) & \psi_{n,m}(2,1)\end{matrix}\right|,
\]
and this is gauge equivalent to \eqref{eq:2ss} if we connect parameters 
$c_j,\rho^+_j,\rho^-_j$ by
\[
\frac{\rho^-_1}{\rho^+_1}=c_1\frac{k_2-k_1}{k_1+k_2},\quad
\frac{\rho^-_2}{\rho^+_1}=c_2\frac{k_1-k_2}{k_1+k_2}.
\]
The $N$-soliton solution is given by the natural extension
\begin{equation}\label{eq:kdvNsol}
f_{n,m}^{[Nss]}=\left|\begin{matrix} 
\psi_{n,m}(1,0) & \psi_{n,m}(1,1)& \cdots &\psi_{n,m}(1,N-1)\\
\psi_{n,m}(2,0) & \psi_{n,m}(2,1)& \cdots &\psi_{n,m}(2,N-1)\\
\vdots & \vdots & \ddots & \vdots \\
\psi_{n,m}(N,0) & \psi_{n,m}(N,1)& \cdots &\psi_{n,m}(N,N-1)\\
\end{matrix}\right|,
\end{equation} 
That this is a solution of \eqref{eq:dHkdvrot} for $\delta$ as given
above can be shown by determinantal manipulations as in  \cite{HZ09}
(cf. Equations (2.20), (5.17b) with bar $\to$ tilde, and (5.18)).

\subsubsection{From bilinear to nonlinear}
Recall that the change of dependent variables from continuous nonlinear to
continuous bilinear by \eqref{eq:nl2bil} involved derivatives, which
are easy. The reverse operation involves integration and is more
involved, especially for the discrete case.

We start with \eqref{eq:dHkdv} and shift it by
$(n,m)\to(n-\frac14,m+\frac14)$ and by $(n,m)\to(n+\frac14,m-\frac14)$
and get \bse\begin{eqnarray*}\label{eq:dHkdvs1}
f_{n+\frac12,m+\frac12}f_{n-1,m\pual}
+\delta^{-1}f_{n,m+1\pual}f_{n-\frac12,m-\frac12}
-(1+\delta^{-1})f_{n-\frac12,m+\frac12}f_{n,m\pual}&=&0,\\ 
f_{n+1,m\pual}f_{n-\frac12,m-\frac12}
+\delta^{-1}f_{n+\frac12,m+\frac12}f_{n,m-1\pual}
-(1+\delta^{-1})f_{n,m\pual}f_{n+\frac12,m-\frac12}&=&0,
\end{eqnarray*}\ese 
respectively. Multiplying the first equation by
$f_{n+\frac12,m-\frac12}$ the second by $f_{n-\frac12,m+\frac12}$ and
subtracting them yields a four term equation and after multiplying it
by
\\ $f_{n,m}/(f_{n+\frac12,m+\frac12}f_{n+\frac12,m-\frac12}
f_{n-\frac12,m+\frac12}f_{n-\frac12,m-\frac12})$
we can write the result as
\begin{align*}
&\frac{f_{n-1,m}f_{n,m}}{f_{n-\frac12,m-\frac12}f_{n-\frac12,m+\frac12}}
-\frac{f_{n+1,m}f_{n,m}}{f_{n+\frac12,m+\frac12}f_{n+\frac12,m-\frac12}}\\
&\hspace{2cm}=\frac1{\delta}
\left(\frac{f_{n,m+1}f_{n,m}}{f_{n+\frac12,m+\frac12}f_{n-\frac12,m+\frac12}}
-\frac{f_{n,m-1}f_{n,m}}{f_{n-\frac12,m-\frac12}f_{n+\frac12,m-\frac12}}\right)
\end{align*}
Now introduce the quantity
\begin{equation}\label{eq:dHWdef}
W:=\frac{f_{n+\frac12,m}f_{n-\frac12,m}}{f_{n,m+\frac12}f_{n,m-\frac12}},
\end{equation}
in terms of which the above equation can be written as
\begin{equation}\label{eq:dHWeq1}
W_{n-\frac12,m}-W_{n+\frac12,m}=\frac1{\delta}\left(\frac1{W_{n,m+\frac12}}-
\frac1{W_{n,m-\frac12}}\right).
\end{equation}
In order to get a convenient form we still make a change in the
discrete variables by
\[
(n+\nu,m+\mu)\mapsto\left(n+m+(\nu+\mu+\tfrac12),
n-m+(\nu-\mu+\tfrac12)\right),
\]
after which equation \eqref{eq:dHWeq1}, when written in terms of
$n':=n+m,\,m':=n-m$, reads 
\begin{equation}\label{eq:Hxkdv}
W_{n',m'}-W_{n'+1,m'+1}=\frac1{\delta}\left(\frac1{W_{n'+1,m'}}
-\frac1{W_{n',m'+1}}\right).
\end{equation}
This now has the standard quad form as it depends on the
corner variables of the quadrilateral as in Figure \ref{F:2}.

If we apply the double continuous limit \eqref{eq:dbl} on the relation
\eqref{eq:dHWdef} we get
\[
\lim_{\epsilon\to 0}\frac1{\epsilon^2}(W-1) = \partial_x^2\log(f(x,t)).
\]
Comparing this to \eqref{eq:nl2bil} we see that equation for $W-1$
should be taken as the discrete KdV equation.

\subsubsection{Relation to the lpKdV version of KdV}
Equation \eqref{eq:Hxkdv} was obtained from the potential KdV equation
\eqref{eq:Hkdvpot} by discretizing its bilinear form
\eqref{eq:Hbilkdv} as \eqref{eq:dHkdvrot}. The discrete bilinear form
was then nonlinearized into \eqref{eq:Hxkdv}. But this equation is
different from the lpKdV equation \eqref{eq:H1} which we announced in the
beginning as being the discrete form of KdV. The reason for the
difference is in that lpKdV \eqref{eq:H1} is discrete {\em potential} KdV
while \eqref{eq:Hxkdv} is discrete KdV or lattice KdV (lKdV).

The explicit relation is obtained as follows: In \eqref{eq:H1} let us
introduce new variables as follows:
\begin{equation}\label{eq:pdH1}
 W_{n,m}:=u_{n,m+1}-u_{n+1,m},\quad Z_{n,m}:=u_{n,m}-u_{n+1,m+1},
\end{equation}
after which \eqref{eq:H1} can be written as
\begin{equation}\label{eq:pdH2}
W_{n,m}\,Z_{n,m}=p^2-q^2.
\end{equation}
According to the definitions \eqref{eq:pdH1} $W,Z$ are related by
\begin{equation}\label{eq:pdH3}
W_{n,m}-W_{n+1,m+1}=Z_{n,m+1}-Z_{n+1,m} 
\end{equation}
and if we solve for $Z_{n,m}$ from \eqref{eq:pdH2} and substitute it into
\eqref{eq:pdH3} we get
\begin{equation}\label{eq:pdH4}
W_{n,m}-W_{n+1,m+1}=(p^2-q^2)\left(\frac1{W_{n,m+1}}-\frac1{W_{n+1,m}}\right),
\end{equation}
which is \eqref{eq:Hxkdv} up to the constant coefficient. The reason
for calling this the lattice KdV and \eqref{eq:H1} the lattice {\em
  potential} KdV is seen from the relation
$W_{n,m}:=u_{n,m+1}-u_{n+1,m}$, which is analogous to $u=v_x$ in  the
continuous case.

\section{Integrability test}
As usual we are mainly concerned with integrable equations, but when
faced with a new equation how can we tell whether it is potentially
integrable? A method that only requires direct computation would be desirable.

In the continuous case we have the Painlev\'e property and for 1D
discrete equations the singularity confinement (SC) idea has been
proposed as an analogous property. SC has turned out to be a very
useful concept as an indicator, although it is only a necessary
condition.

For discrete equations there is also the concept of ``algebraic
entropy'' which states that the complexity of the iterates should not
grow exponentially, but only polynomially. Complexity is here measured
by the degree of an iterate with respect to the initial values. This
is computationally demanding but can be automatized.

As an example consider the corner initial values as in Figure
\ref{F:3} a) and a quadratic equation (such as \eqref{eq:H1}). If we
define $u_{n,0}=x_n,\,u_{0,m}=y_n\, (y_0=x_0)$ we find that
generically $u_{1,1}$ is quadratic over linear in the initial values,
$u_{1,2}$ and $u_{2,1}$ are cubic over quadratic, and $u_{2,2}$ is
order six over order five. However, for the particular case of
\eqref{eq:H1}, which is integrable, the numerator and denominator of
$u_{2,2}$ have a common factor which can be canceled and the degrees
are just order five over order four. Such cancellations continue for
higher orders. According to \cite{TGR01}, for \eqref{eq:H1} the degree
of the numerator is $d_{n,m}=nm+1$, that is the growth is
polynomial. For a generic quadratic equation the asymptotic growth of
degrees is exponential.

\section{Summary}
In this brief introduction to the discrete or lattice soliton
equations we have looked at some of their important features, and as
an example we have given explicit details for the Korteweg-de Vries
equation. We have discussed how the permutation property of B\"acklund
transformations can be interpreted as discrete equations on a $\mathbb
Z^2$ lattice and how lattice evolutions are typically defined. The
fundamental idea of hierarchy of equations is in the lattice setting
provided by multidimensionality. The multidimensional consistency
condition was discussed in detail, along with its consequence of
providing Lax pair and B\"acklund transformation, which was used to
construct a one-soliton solution. We have also discussed continuum
limits and discretization, in particular Hirota's discretization and
the ensuing soliton solutions.

For further introductory material we refer the reader to the book \cite{HJN}.

\section*{Acknowledgements}
I would like to thank Da-jun Zhang for comments on the manuscript.

\end{document}